\documentclass{ws-procs961x669} 
\begin{document}
\title{Dynamics of relativistic electrons in non-uniform magnetic fields and its applications in quantum computing and astrophysics}
\author{Srishty Aggarwal$^*$ and Banibrata Mukhopadhyay}
\address{Department of Physics, Indian Institute of Science, Bengalore, Karnataka 560012, India\\
$^*$srishtya@iisc.ac.in}

\begin{abstract}
    We explore the two-dimensional motion of relativistic electrons when they are trapped in magnetic fields having spatial power-law variation. Its impacts include lifting of degeneracy that emerged in the case of the constant magnetic field, special alignment of Landau levels of spin-up and spin-down electrons depending on whether the magnetic field is increasing or decreasing from the centre, splitting of Landau levels of electrons with zero angular momentum from that of positive one and the change in the equation of state of matter.  Landau quantization (LQ) in variable magnetic fields has interdisciplinary applications in a variety of disciplines ranging from condensed matter to quantum information. As examples, we discuss the increase in quantum speed of the electron in presence of spatially increasing magnetic field; and the attainment of super Chandrasekhar mass of white dwarfs by taking into account LQ and Lorentz force simultaneously.
\end{abstract}

\bodymatter

\section{Introduction}
Landau Quantization (LQ) is the phenomenon of the quantization of the cyclotron orbits of charged particles in presence of magnetic field. LQ has been widely discussed for uniform magnetic fields both in non-relativistic \cite{1965qume.book.....L} as well as relativistic \cite{1998rqm..book.....S} cases. It leads to many interesting effects like quantum Hall effect, de Has Van Alphen effect, Shubnikov oscillations, modification to the equation of state (EOS) and change in the neutron drip line. The modified EOS was later proved to be useful in explaining the mass of super Chandrasekhar white dwarfs \cite{PhysRevD.86.042001}. 

However, a complete uniform magnetic field is an idealistic realization. What if it is non-uniform as in astrophysical systems and plasmas? In white dwarfs, neutron stars and even main sequence stars, the magnetic field varies drastically from centre to surface. Plasma can act as a source of both increasing and decreasing magnetic fields depending on its properties. Even in laboratories, fluctuations in the magnetic field are a common sight. This motivates us to initiate the exploration of the dynamics of a relativistic electron in presence of strictly spatially variable magnetic fields \cite{Aggarwal21}.

We show different arrangements of the energy levels of an electron when it is trapped in variable magnetic fields. Further, we explore its applications in the fields of quantum information by showing an increase in quantum speed of an electron in presence of a spatially increasing magnetic field, and astrophysics, in the magnetized white dwarfs.

\section{Solution of Dirac equation in presence of magnetic fields}
The Dirac equation in the presence of magnetic field for an electron of mass $m_{e}$ and charge $q~(-e)$ is given by
\begin{equation}
i\hbar\frac{\partial\Psi}{\partial t} = \left[ c\boldsymbol{\alpha}\cdot\left(-i\hbar\textbf{$\nabla$}-\frac{q\textbf{A}}{c}\right) + \beta m_{e}c^2\right]\Psi,
\label{eq1}
\end{equation}
where $\boldsymbol{\alpha}$ and $\beta$ are Dirac matrices, $\textbf{A}$ 
is the vector potential, $\hbar$ is the reduced Planck constant and 
$c$ is the speed of light. For stationary states, we can write
\begin{equation}
	\Psi = e^{-i\frac{Et}{\hbar}}\begin{bmatrix}
  \chi \\
 \phi \\
 \end{bmatrix}
\label{matrix},
 \end{equation} 
where $\phi$ and $\chi$ are 2-component spinors.

Therefore, the decoupled equation for $\chi$ obtained from Eq. (\ref{eq1}) is 
\begin{equation}
(E^2-m_{e}^2c^4)\chi = \left[c^2\left(\pi ^2 - \frac{q\hbar}{c}\boldsymbol\sigma\cdot \textbf{B}\right)\right]\chi, 
\label{eq3}
\end{equation}
where $\boldsymbol{\pi}=-i\hbar\textbf{$\nabla$}-q\textbf{A}/c$. 

We choose a simple 
power law variation of the magnetic field in compliance with Maxwell's equations as
\begin{equation}
\textbf{B} = B_{0}\rho^n \hat{z},
\end{equation}
in cylindrical coordinates $(\rho,\phi,z)$. Such a field profile can take into account the uniformity in magnetic field when $n=0$ as well as an increase and a decrease in fields with respect to radial coordinate with $n>0$ and $n<0$ respectively, satisfying other physics intact.
Using a gauge freedom for the vector potential \textbf{A}, we choose
\begin{equation}
 \textbf{A}=B_{0}\frac{\rho^{n+1}}{n+2} \hat{\phi} = A\hat{\phi}.
\end{equation}
Let
\begin{equation}
\chi = e^{i\left(m\phi+\frac{p_{z}}{\hbar}z\right)}R_{\pm}(\rho).
\label{eq4}
\end{equation}
Since the electron is confined to a plane perpendicular to the direction of magnetic field ($z$-direction), $p_z=0$. By substituting $\chi$ from Eq. (\ref{eq4}) into Eq. (\ref{eq3}), and dividing it by $m_{e}^2 c^2$, Eq. (\ref{eq3}) becomes
\begin{equation}
    \alpha R_{\pm}(\rho)  = -\lambda_{e}^2\left[\frac{\partial ^2}{\partial \rho^2}+\frac{1}{\rho}\frac{\partial}{\partial\rho}-\frac{m^2}{\rho^2}\right]\tilde{R}_{\pm}  + \left[\left(\frac{kB_0\rho^{n+1}}{n+2}\right)^2 + k\lambda_{e}\left(-\frac{2m}{n+2}\pm 1\right)B_0\rho^n\right]R_{\pm},
\label{eq6}
\end{equation}
where $\alpha$ is the eigenvalue of the problem and is equal to $\epsilon^2-1$, $\epsilon=E/m_e c^2$, $\lambda_e = \hbar/m_e c$ and $k=e/m_e c^2$.
Further, choosing $R_{\pm}(\rho)= u_{\pm}(\rho)/\sqrt{\rho}$, the above equation is reduced to
\begin{equation}
    \alpha u_\pm = \left(-\lambda_{e}^2\frac{\partial ^2}{\partial \rho^2}+V_{eff}\right)u_\pm,
\label{eq20}
\end{equation}
such that
\begin{equation}
    V_{eff} = -\lambda_{e}^2\left[\frac{1}{4\rho^2}-\frac{m^2}{\rho^2}\right] + \left(\frac{kB_0\rho^{n+1}}{n+2}\right)^2+ k\lambda_{e}\left(-\frac{2m}{n+2}\pm 1\right)B_0\rho^n.
\end{equation}

\begin{figure}
    \centering
    \includegraphics[scale=0.7]{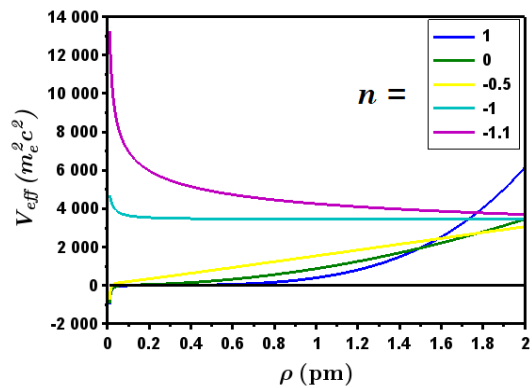}
    \caption{The variation of effective potential for different $n$ for
$B_0 = 10^{15}$ G pm$^{-n}$. The black horizontal line represents 
$V_{eff}=0$. Various potentials at $\rho=1$~pm from bottom to top represent
	 $n=1,0,-0.5,-1\: {\rm and}-1.1$ respectively.}
    \label{fig:veff}
\end{figure}
The variation of $V_{eff}$ for different $n$ is shown in Fig. \ref{fig:veff}. It can be seen that
$V_{eff}$ is completely repulsive for $n$ less than or equal to -1, whose solution will depend on the distance from the
source (origin of the system) upto which a particle can move. Therefore, the energy
eigenvalues for such cases depend upon where we put a hard wall making the system
equivalent to confining the electron in a box. However, we do not want to apply any such
restrictions on the electron. We, therefore, limit our analysis to the cases with $n$ greater than -1. 

We use "Shooting and Matching' method to solve Eq. (\ref{eq20}). For computational analysis, we choose $\rho$ and $B_0$ in the units of picometer (pm) and G $\rm pm^{-n}$ respectively.

For uniform magnetic field, i.e. $n=0$, the eigenvalue of level $\nu$ is given by
\begin{equation}
\alpha_{\nu} = 2k\lambda_{e}B_{0}\left (\nu+\frac{|m|}{2}-\frac{m}{2}+\frac{1}{2}\pm\frac{1}{2}\right),
\label{eq12}
\end{equation}
where `$m$' is the azimuthal quantum number \cite{1965qume.book.....L}.
One can easily see from Eq. (\ref{eq12}) that ground state eigenvalue 
($\alpha_{0}$) is 0 and all the other energy levels are doubly degenerate. 
Also, energies are same for $m\ge 0$, which
turn out to be 
\begin{equation}
E^2=p_z^2c^2+m_e^2c^4\left(1+2\nu\frac{B_0}{B_c}\right),
\label{Ev}
\end{equation}
where $B_c=m_e^2c^3/e\hbar$, the Schwinger limit of pair production.

\section{Eigenspectra and dispersion relations}
\subsection{$m=0$}

\begin{figure}
    \centering
    \includegraphics[scale=0.3]{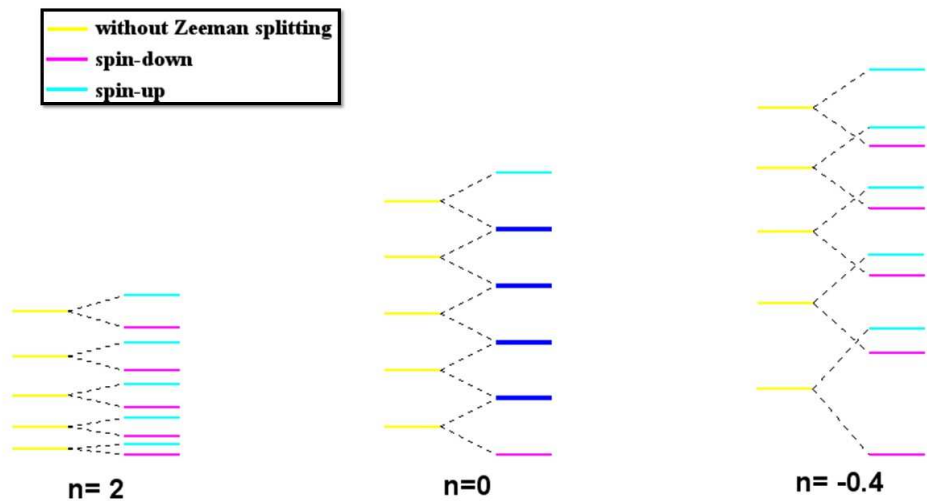}
    \caption{Comparison of eigenlevels and the splitted states of $+\sigma.B\:{\rm and}\:-\sigma.B$ for $n=2,0$ and $-0.4$.}
    \label{fig:pnspec}
\end{figure}

\begin{figure}
    \centering
    \includegraphics[scale=0.4]{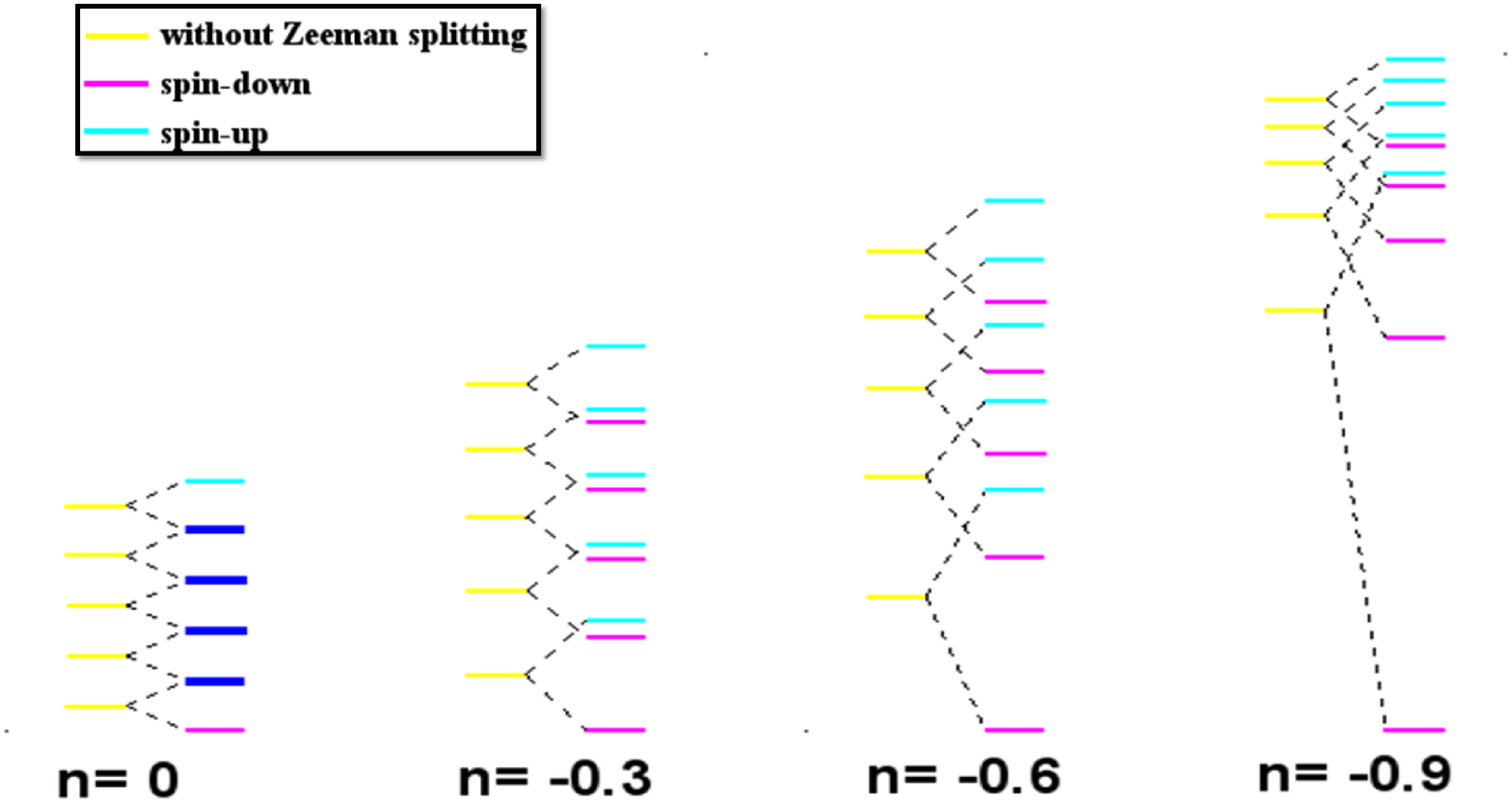}
    \caption{Comparison of eigenlevels and the splitted states of $+\sigma.B\:{\rm and}\:-\sigma.B$ for $n=0$ and when $n$ is negative.}
    \label{fig:nspec}
\end{figure}

Figures \ref{fig:pnspec} and \ref{fig:nspec} show the eigenspectra for different $n$, obtained from Eq. (\ref{eq6}) by taking $m=0$. On comparison of the eigenlevels without Zeeman effect (yellow-solid lines in the left) in Fig. \ref{fig:pnspec}, it can be observed that the spacing between the eigenlevels increases for $n>0$, remains same if $n=0$ and decreases for $n<0$, from ground to higher levels. It owes to the chosen magnetic field variation in the region. If the magnetic field is spatially increasing, it has low strength near the centre and becomes strong near the outer boundary. Hence, the lower levels have lesser spacing, and the gap between consecutive levels increases for higher levels. The opposite is true for a decaying field profile.

Breaking of degeneracy that arose in uniform magnetic field takes place for both increasing and decreasing magnetic fields if Zeeman effect is included. However, the alignment of levels for spin-up and spin-down electrons is quite different for these two cases as shown in Fig. \ref{fig:pnspec}. While the spin pattern for $n>0$ is \textit{dududu...}, it becomes \textit{ddudud...} for $n<0$ until $n=-0.6$, as can be inferred from Fig. \ref{fig:nspec}. Below -0.6, the spin pattern is highly ambiguous, say for $n=-0.8$, it is \textit{dddudud...}, whereas for $n=-0.9$, it changes to \textit{ddddudud...}. Here, \textit{d} and \textit{u} denote the levels for spin-down and spin-up electrons respectively. It is remarkable that the ground level for the spin-down electron always lies at 0 for all $n$.

One can, thus, obtain the desired spacing between the eigenlevels using non-uniform magnetic fields. For example, we can prepare a system
with electrons confined to lower levels in a decaying magnetic field, say with $n=-0.8$ or $-0.9$. Such a system will then have only spin-down electrons due to the availability of lower levels for the same. Thus, we can achieve a single spin system using such variable magnetic fields. Also, a spin-based transition can be observed in presence of increasing magnetic fields, if we restrict the energy of electron between spin-up ground level and spin-down first excited state energies.
\begin{table}[htbp]
    \centering
    \tbl{The values of the constants of Eq. (\ref{eq alpha}) for various $n$. Here $B_0$ in Eq. (\ref{eq alpha}) is chosen in the units of $10^{15} \:$ G pm$^{-n}$}{
    \begin{tabular}{ccc}
    \hline
       \:\,$n$\:\,  & $\:\,C_3\:\,$ & $\:\,C_5\:\,$ \\
       \hline
      -0.5 &  195.66 &  0.484 \\
      -0.4 &  134.63 & 0.486 \\
      -0.3 &  97 & 0.488 \\
      -0.2 &  72.5 & 0.4934 \\
      -0.1 &  56 & 0.50 \\
      0 & 44.42 & 0.50 \\
      1 & 10.95 & 0.5156 \\
      2 &  5.72 & 0.50 \\
      3 &  3.965 & 0.51 \\
      4 &  3.15 & 0.51 \\
      \hline
    \end{tabular}}
    
    \label{tab:constants}
\end{table}

We obtain a general expression for the eigenvalue of a level $\nu$ for arbitrary $n$ using trial and error method, given by \cite{Aggarwal21}
\begin{equation}
\alpha_{\nu\pm} = C_3\: B_0^{\frac{2}{n+2}}\:(\nu+C_5)^{\frac{2+2n}{n+2}}\left[1\pm \frac{C_5}{ (\nu+C_5)}\right].
\label{eq alpha}
\end{equation}
where $C_3$ and $C_5$ are constants whose value depends on $n$. The values of these constants for some $n$ are given in Table \ref{tab:constants}.

\subsection{$m\neq0$}

\begin{figure}
    \centering
    \includegraphics[scale=0.32]{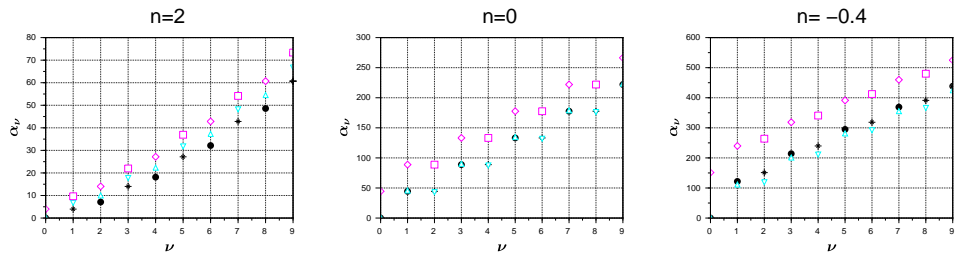}
    \caption{The variation of eigenvalue with the eigen-index for $m = 0$
with $-\sigma.B$ (black solid circles) and +$\sigma.B$ (black asterisks);
$m=+1$ with $-\sigma.B$ (cyan upward triangles) and +$\sigma.B$ (cyan
downward triangles); $m=-1$ with $-\sigma.B$ (magenta diamonds) and 
 $+\sigma.B$ (magenta squares), for $n=2,0$ and $-0.4$.}
    \label{fig:mspec}
\end{figure}

We compare the eigenvalues between positive $m$ ($m= 1$) and negative $m$ ($m= -1$) with $m=0$ for different variations in the magnetic fields. When $n=0$, positive $m$ does not have any impact on the eigenvalues in comparison with $m= 0$, while negative $m$ leads to the increment in energies, as depicted in Fig. \ref{fig:mspec} as well as can be inferred from Eq. (\ref{eq12}). The Fig. \ref{fig:mspec} shows that for variable magnetic fields, the trend with respect to negative $m$ remains same as that for the uniform one, however, the impact of positive $m$ changes for the varying field. While it leads to an increase in energies for $n>0$, the energies decrease, if $n$ is negative. 

%The above behavior could be an effect arising due to the rotation of 
%particles and the field variation. Generally, rotation opposite to the direction of the magnetic field is difficult
%and in the same direction, it is easy.  
%While a positive $m$ indicates the former, a negative $m$ implies the latter. When the magnetic field is homogeneous, a rotation in the direction to the magnetic field is not expected to require any extra force when there is no change of field magnitude (no force due to magnetic pressure), leading to no change in energy. However, to rotate a particle opposite to its natural direction, extra force is required, hence raising the energies in uniform as well as non-uniform magnetic fields.\textbf{How to explain the behavior for postive n} (for negative n) In an non-homogeneous field, as the magnitude of field changes, the particle has to overcome the force due to magnetic pressure, even if the direction is same as of the magnetic field. Hence, its own energy dissipates, leading to the less energy to align with the magnetic field.

\section{Finding ground state energy using variational method}

We can find the upper bound of the ground state eigenvalue $\alpha_0$ using the variational principle \cite{Griffith}, which states that
\begin{equation}
    E_{gs} \leq \langle H\rangle,
    \label{eq:var}
\end{equation}
where $E_{gs}$ represents the ground state energy and $\langle H\rangle$ is the expectation value of Hamiltonian.

We choose the ground state wavefunction for both the spins of electron as
\begin{equation}
    R_{0}  = e^{-\frac{KB_0\rho^{n+1}}{\lambda_e(n+2)^2}}.
\end{equation}
The normalisation constant ($N_0$) is given by
\begin{equation}
    N_0  = \sqrt{\frac{(n+2)}{\Gamma\left(\frac{2}{n+2}\right)}\left(\frac{2KB_0}{\lambda_e(n+2)^2}\right)^{\frac{2}{n+2}}},
\end{equation}
where 
\begin{equation}
    \Gamma \left(\frac{a}{b}\right) = b \int_0^{\infty} x^{a-1}e^{-x^b}dx. \nonumber
\end{equation}
Hence, the normalised wavefunction $\tilde{R}_0 = N_0 R_0$. Therefore,
\begin{equation}
\tilde{R}'_{0} = -\rho^{n+1}\frac{KB_0}{\lambda_e(n+2)}\tilde{R}_0,
\end{equation}
and
\begin{equation}
\tilde{R}''_{0} = -(n+1)\rho^n\frac{KB_0}{\lambda_e(n+2)}\tilde{R}_0 + \left(\frac{KB_0\rho^{n+1}}{\lambda_e(n+2)}\right)^2\tilde{R}_0.
\end{equation}
Hence,
\begin{eqnarray}
&&\alpha_{0_{Th}\pm} = \langle \tilde{R}_0|\:\alpha_{\pm}\:|\tilde{R}_0\rangle \label{Leq8}\\
&& \quad \:\;= \frac{KB_0\lambda_e (1\pm1)}{\Gamma\left(\frac{2}{n+2}\right)}\left(\frac{2KB_0}{\lambda_e(n+2)^2}\right)^{-\frac{n}{n+2}}.
\label{eq9}
\end{eqnarray}
Thus, Eq. (\ref{eq9}) provides an upper bound on the ground state energies for spin-down and spin-up electrons. (Eq. \ref{eq9} can be obtained from Eq. \ref{eq6}). 

\begin{table}
    \centering
     \tbl{Comparison of analytical and computational 
     $\alpha_0$ values
     at different $n$ for $B_0=10^{15}~\rm{G}~pm^{-n}$.}
    {\begin{tabular}{ccc}
    \hline
    \:\,n\:\, & $\:\,\alpha_{0_{ Comp}+}\:\,$     & $\:\,\alpha_{0_{ Th}+}\:\,$ \\
    \hline
    -0.5 & 225.83 &  256.70084 \\
    -0.4 & 151.124 &  162.464 \\
    -0.3 & 105.612 &  109.563 \\
    -0.2 & 76.716 & 77.840 \\
    -0.1 & 57.55 & 57.746 \\
    0 & 44.4 &  44.418 \\
    1 & 8.702 & 10.090 \\
    2 &  3.969 & 5.700  \\
    3 & 2.534 &   4.427 \\
    4 & 1.907 &  3.953 \\
    \hline
    \end{tabular}}
   
    \label{tab:alpha0}
\end{table} 
Note that $\alpha_{0_{Th}-}$ is zero for all $n$, which is the exact eigenvalue for the ground state of spin-down electron. For spin-up electron, the above expression provides the exact ground state eigenvalue only for the uniform magnetic field, but deviates for the non-uniform one.
The comparison of actual $\alpha_0$ obtained computationally ($\alpha_{0_{Comp}+}$) and the approximate analytical one obtained in this section ($\alpha_{0_{Th}+}$) for spin-up electron at $B_0=10^{15}\:\rm G\:pm^{-n}$ for different $n$ is shown in Table \ref{tab:alpha0}. The deviation between the two values increases as one goes away from $n=0$ in either direction. It is noteworthy that the power-law dependence of $\alpha_{0_{Th}}$ on $B_0$ in Eq. (\ref{eq9}) is ${2/(n+2)}$ which is same as in Eq. (\ref{eq alpha}), where $\alpha$ for a general $\nu$ is obtained using an appropriate ansatz. 

\section{Applications}

In general, LQ in non-uniform magnetic fields can be useful in multiple domains of physics ranging from quantum information to condensed matter systems. Here, we will discuss its applications in quantum information via the increase in quantum speed of an electron and in astrophysics for the super-Chandrasekhar white dwarfs.  

\subsection{Increase of quantum speed}
Quantum speed of a particle is defined as the speed of its transition from one energy level to the other. It has a direct influence on the processing speed of quantum information. The electron levels are equivalent to qubits in quantum information. Thus, if the transition between the levels takes place at higher pace, switching between the qubits will be more rapid, thereby, leading to faster processing of quantum information. Therefore, attaining higher quantum speed is one of the major requirements of the researchers in the field of quantum computation.

Assuming that the initial state of the electron is the superposition of the two consecutive states: the ground and the first excited one, given by
\begin{equation}
 \Psi(\rho,0) = \frac{1}{\sqrt{2}}\left[\psi_0(\rho)+\psi_1(\rho)\right],
 \end{equation} 
with $m=0$, $p_z=0$ and the respective energies $E_0$ and $E_1$.
For the transition of an electron from ground state to first excited state, the minimum time of evolution of wavefunction ($T_{min}$) is evaluated using Mandelstam-Tamm bound \cite{mandelstam1945energy} as
\begin{equation}
 T_{min} = \frac{\pi\hbar}{2\Delta H},
 \end{equation}
 where
 \begin{equation}
 \Delta H = \frac{E_1 - E_0}{2}.
 \end{equation}
 The radial displacement of electron in time $T_{min}$ is given by
\begin{equation}
  \rho_{disp}  =2\left\vert\int^\infty_0 \rho D_S(\rho)d\rho\: \right\vert,   
\end{equation}
where
\begin{equation}
D_s(\rho) = \psi^{\dagger}_{0}\:\rho\:\psi_{1}.
\label{eq6q}
\end{equation}
Hence, the quantum speed of electron is
\begin{equation}
\tilde{\rm v} = \frac{\rho_{disp}}{T_{min}},
\end{equation}

\begin{figure}
\centering
\includegraphics[scale=.4]{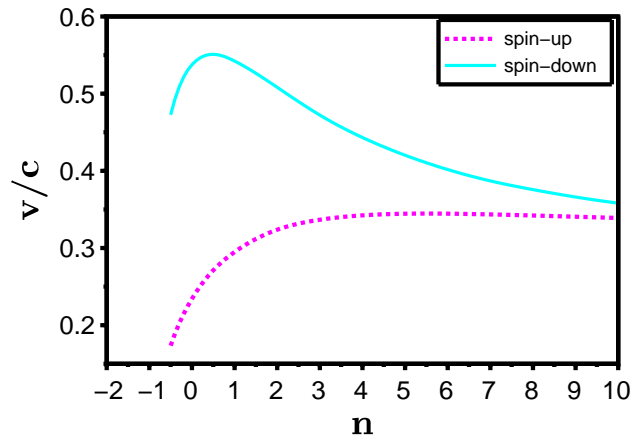}
\caption{Variation of quantum speed of spin-up and spin-down electrons for transition from 
	ground state to first excited state with different $n$ at $B_0 = 10^{16}\:{\rm{G}\:pm}^{-n}$.}
\label{qsl}
\end{figure}

Figure \ref{qsl} illustrates the variation of the quantum speed of an electron with the change in non-uniformity index of the magnetic field in the relativistic regime.  
As it can be seen from the figure, the quantum speed of the electron increases with increasing $n$; reaches the maximum and then begins to decrease. Whereas quantum speed is maximum for $n=1$ for spin-down electron, $n$ has to be 6 to reach the highest speed for spin-up electron for $B_0=10^{16}\:\rm G\:pm^{-n}$. This is associated with the varied alignment of 
energy levels in non-uniform magnetic fields lifting the degeneracy between $n<0$ and $n>0$ as shown in 
Fig. \ref{fig:pnspec}. Therefore, if a spin-up electron is trapped in a magnetic field that is spatially increasing in 
magnitude even linearly ($n=1$), then we can achieve
a higher speed of transition of the electron as shown in Fig. \ref{qsl}.

\subsection{Super-Chandrasekhar white dwarfs}
Astrophysical bodies like white dwarfs, neutron stars and magnetars are the natural sites to observe non-uniform magnetic fields, wherein, the difference between the central and the surface magnetic fields lies upto $3-4$ orders of magnitude or even higher. 

The main impact of LQ in stellar physics is to modify the underlying EOS due to the discretization of energy levels in presence of strong magnetic fields. For a constant magnetic field, the energy levels are degenerate and the spacing between the eigenlevels is constant, but for a variable magnetic field case, the degeneracy breaks down and the spacing between levels is non-uniform, as shown in Figs. \ref{fig:pnspec} and \ref{fig:nspec}, which brings a considerable change in EOS in presence of non-uniform magnetic fields.

\begin{figure}
    \centering
    \includegraphics[scale=0.3]{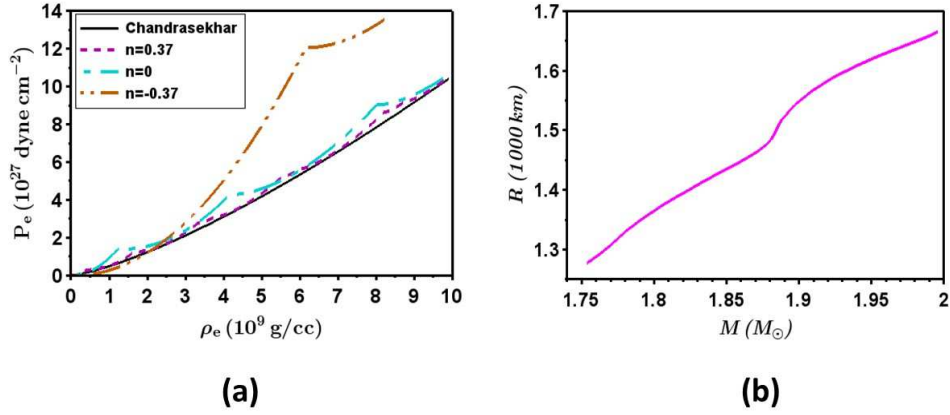}
    \caption{(a) EOS for different $n$ at $\epsilon_{Fmax} = 17$ for $B_0 = 2\times 10^{15}\: \rm G\:pm^{-n}$. (b) Mass-radius relation of magnetized white dwarf for the magnetic field profile chosen in Eq. (\ref{bprof}).}
    \label{fig:mreos}
\end{figure}
In a non-uniform magnetic field, the electron number density at zero temperature is given by
\begin{eqnarray}
	\nonumber
	%n_e &=& \\ \nonumber
	%{n_e}_+ + {n_e}_- \\ \nonumber 
	n_e= \frac{1}{(2\pi)^2\lambda_e^3}\left(\sum_{\nu=0}^{\nu=\nu_{m-}}\beta _-(\nu)x_{F-}(\nu)+\sum_{\nu=0}^{\nu=\nu_{m+}}\beta _+(\nu)x_{F+}(\nu)\right),\\
\label{eq35}
\end{eqnarray}
where
`$+$' sign and `$-$' sign indicate spin-up and spin-down respectively, 
%$x_{F}=p_F/m_e c$,
\begin{equation}
x_{F\pm}(\nu) = \left[\epsilon_F^2 - (1+\alpha_{\pm}(\nu)\right]^{\frac{1}{2}}
\label{eq36}
\end{equation}
and
\begin{equation}
\beta_{\pm} = \left(\alpha_{\pm}(\nu+1) - \alpha_{\pm}(\nu-1)\right)/2.
\end{equation}
Note that $\nu_{m_{\pm}}$ is the largest integer value of $\nu$ such that 
\begin{equation}
    \alpha_{\nu_{\pm}}\le\epsilon_F^2-1.
\end{equation}

The pressure of an electron gas at zero temperature is given by
\begin{eqnarray}
%\begin{align}
	\nonumber
	&&P_e=\frac{m_ec^2}{(2\pi)^2\lambda_e^3}\left(\sum_{\nu=0}^{\nu=\nu_{m-}}\beta_-(\nu)(1+\alpha_-(\nu))f_2\left[\frac{x_{F-}(\nu)}{(1+\alpha_-(\nu))^{1/2}}\right] \right. \\ &&+\left.
\sum_{\nu=0}^{\nu=\nu_{m+}}\beta_+(\nu)(1+\alpha_+(\nu))f_2\left[\frac{x_{F+}(\nu)}{(1+\alpha_+(\nu))^{1/2}}\right]\right),
\label{eq39}
%\end{align}
\end{eqnarray}
where
\begin{equation}
f_2(z) = \frac{1}{2}\left(z\sqrt{1+z^2} - ln (z+\sqrt{1+z^2})\right).
\end{equation}

Figure \ref{fig:mreos}(a) shows how the EOS changes with $n$ at $B_0=2\times 10^{15}\:\rm G\:pm^{-n}$. For a given $\epsilon_{Fmax}$, the allowed number of levels decreases with decreasing $n$. With the decrease in number of levels, EOS becomes stiffer and softer in high and low density regimes respectively. Although, the spatially increasing magnetic field is not common in stellar objects, we also show the change in EOS of electron in presence of an increasing magnetic field that may come handy for future explorations.

To probe the effect of LQ and Lorentz force simultaneously in white dwarfs, we choose the following sample magnetic field profile.
\begin{align}
	B = 
	\begin{cases}
	B_0\hat{z}, & {\rm if}\,\,\,\rho < 850~{\rm km},\\
	B_0\left(\frac{\rho}{\rm 1~km}\right)^{-0.37}\hat{z}, & {\rm if}\,\,\,850~{\rm km}\leq \rho \leq 900~{\rm km},\\
	B_0\left(\frac{\rho}{\rm 1~km}\right)^{-0.99}\hat{z}, & {\rm otherwise},
	\end{cases}
	\label{bprof2}
\end{align}
such that the central and the surface magnetic field are $2\times10^{15}$ G and around $10^{12}$ G respectively.
Also, $(\textbf{B}\cdot\nabla)\textbf{B}=0$ and $\nabla\cdot\textbf{B}=0$ for the chosen profile. Therefore, the non-rotating white dwarfs will be spherical in shape. 
Hence, in spherical polar coordinates with $\theta=\pi/2$, the field 
profile is given by
\begin{align}
	B = 
	\begin{cases}
    -B_0\hat{\theta}, & {\rm if}\,\,\,r < 850~{\rm km},\\
	-B_0\left(\frac{r}{\rm 1~km}\right)^{-0.37}\hat{\theta}, & {\rm if}\,\,\,850~{\rm km}\leq r \leq 900~{\rm km},\\
	-B_0\left(\frac{r}{\rm 1~km}\right)^{-0.99}\hat{\theta}, & {\rm otherwise},
	\end{cases}
	\label{bprof}
\end{align}

The mass and radius of a white dwarf can be obtained by solving 
\begin{equation}
	\frac{d}{dr}\left(P_e+\frac{B^2}{8\pi}\right)=-\frac{GM(r)
	(\rho_e+\rho_B)}{r^2},
	\label{msbal}
\end{equation}
\begin{equation}
	\frac{dM(r)}{dr}=4\pi r^2(\rho_e+\rho_B),
\end{equation}
where $\rho_B$ is the magnetic density, $B^2=\textbf{B}\cdot\textbf{B}$, $\rho_e=n_e m_p\mu_e$, $m_p$ is the 
mass of proton, $\mu_e$ is the mean molecular weight per electron and
$G$ is Newton's gravitation constant. 

The mass-radius relation, shown in Fig. \ref{fig:mreos}(b), depicts the existence of super-Chandrasekhar mass white dwarfs in presence of central strong magnetic field. Although, here, the surface magnetic field obtained is around $10^{12}$~G, the results are unaffected even if the surface magnetic field of white dwarfs is lower, say around $10^9$~G which is detectable.

Here, for $r<850$~km, since, magnetic field is uniform, EoS would be Landau quantized for the same magnetic field at 
$B_0=2\times 10^{15}$ G. For $r\ge 850$~km, the field decays to a lower strength so that Chandrasekhar's
non-magnetic EoS suffices. Only at the interface around 850~km, non-uniform field based LQ applies in EoS,
but in a very tiny region. Hence practically, for the present example, LQ EOS based on non-uniform field does not play an important role in controlling the dynamics of magnetised white dwarf. However, it helps to account for the whole system including the small zone around 850~km that contains the decay of the magnetic field.

%Another possible application could be in the field of astrophysics where the decaying magnetic field from centre to surface is a common sight. However, for the discussed LQ in variable magnetic field, for $B_0~10^{15}\:G\:pm^{-n}$, the variation in magnetic field has to be in picometer scale, which indicates a rapid decay of field.  
\section{Summary}
LQ in strictly spatially variable magnetic fields is a new venture on its own. It leads to the different alignment of energy levels of an electron as well as the breaking of spin degeneracy, depending on the variation of magnetic field. Hence, it can be useful in the multitude of fields wherein one can attain the desired spacing and alignment of levels through an appropriate non-uniformity in magnetic fields. 

We have explored its application in quantum computing through an increase in quantum speed of electron in spatially increasing magnetic field. We also discuss the increase in stiffness and softenss of EOS at high and low densities in presence of non-uniform magnetic fields as compared to the case of uniform magnetic field (as well as Chandrasekhar EOS). This, we, further, use to understand the properties of magnetized white dwarfs in the small zone of rapidly decaying magnetic field.

\bibliographystyle{ws-procs961x669}
\bibliography{bibliography}

%I have not cited my own paper anywhere.
\end{document}